\documentclass[prb,a4paper,twocolumn,showpacs,superscriptaddress,amssymb]{revtex4}
\pdfoutput=1
\usepackage{graphicx}
\usepackage{hyperref}
\usepackage{amssymb}
\usepackage{mathptmx}      

\begin{document}

\title{Universal Onset of Quantum Turbulence in Oscillating Flows \\ and Crossover to Steady Flows}

\author{R. H\"anninen}
\affiliation{Low Temperature Laboratory, Helsinki University of Technology, FIN-02015 TKK, Finland} 
\author{W. Schoepe}
\affiliation{Institut f\"ur Experimentelle und Angewandte Physik, Universit\"at Regensburg, D-93040 Regensburg, Germany} 
\date{\today}

\begin{abstract}  
The critical velocity $v_c$ for the onset of quantum turbulence in oscillatory flows of superfluid helium is universal and scales as $v_c\sim\sqrt{\kappa\omega}$, where $\kappa$ is the circulation quantum and $\omega$ is the oscillation frequency. This result can be derived from a general argument based on the ``superfluid Reynolds number''. Only the numerical prefactor may depend somewhat on the geometry of the oscillating object because the flow velocity at the surface of the object may differ from the velocity amplitude of the body. A more detailed analysis derived from the dynamics of the turbulent state gives $v_c  \approx \sqrt{8\kappa \,\omega/\beta}$, where $\beta\sim1$ depends on the mutual friction parameters. This universality is compared with the recently discovered universality of classical oscillatory flows. We also discuss the effect of remanent vorticity on the onset of quantum turbulence. Finally, by employing the ``superfluid Reynolds number'' again, we argue how $v_c$ changes when the steady case $\omega = 0$ is approached. In that case $v_c$ scales as $\kappa/R$, where $R$ is the size of the object.  


\end{abstract}

\pacs{67.25.dk, 67.25.dg, 47.27.Cn}

\maketitle

\section{Introduction}\label{intro}

The critical velocity $v_c$ for the onset of quantum turbulence caused by a macroscopic body oscillating in superfluid helium is discussed. Experiments with spheres \cite{archiv} and vibrating wires \cite{yano,pickett} give evidence for a universal scaling property of oscillatory superflows, namely $v_c\sim\sqrt{\kappa\omega}$, independent of the particular geometry of the vibrating object that drives the oscillating superflow. This experimental result can be derived from the ``superfluid Reynolds number'' $Re_s = vl/\kappa$, where $l$ is some characteristic length scale \cite{sRe}. Turbulence may be expected when $Re_s \ge 1$. At oscillation amplitudes $a = v/\omega$ that are small compared to the size $R$ of the object, $a \ll R$, the amplitude $a$ is the characteristic length scale while $R$ may be considered to be infinite. This leads to the above scaling \cite{jltp}. Interestingly, for classical oscillatory flows a universal scaling has recently been discovered \cite{Ekinci} which might be related to our case. 

More detailed information can be obtained from the dynamical behaviour of the turbulent vortex tangle as recently discussed by Kopnin \cite{kolya} for the case of counterflow turbulence at {\it constant} velocity. Extending this scenario qualitatively to the case of {\it oscillating} flows gives \cite{jltp} $v_c  \approx \sqrt{8\kappa \,\omega/\beta}$, where $\beta = A(1-\alpha^{\prime})-B\alpha$ is derived from the mutual friction parameters $\alpha^{\prime}$ and $\alpha$ with coefficients $A,B\sim$1 (more precise values for $A$ and $B$ would require numerical simulations with some particular geometry). Below 1 K where mutual friction is small in $^4$He ($\alpha^{\prime}, \alpha \approx 0$) we can set $\beta =1$ (assuming $A,B=1$) while towards higher temperatures $\beta$ slowly decreases to 0.79 at 1.9 K (the highest temperature where $v_c$ could be resolved with spheres). This leads to a slight increase of $v_c$ with temperature that is in good agreement with the data on spheres \cite{jltp}. Very close to the Lambda transition $\beta$ might decrease very fast and ultimately become negative. This would imply that because of large mutual friction superfluid turbulence cannot exist anymore, analogous to the situation in superfluid $^3$He above $T/T_c \approx 0.6$ \cite{Rota}. 

The numerical prefactor $\sqrt{8} = 2.83$ is only an estimate because the model is still qualitative as long as there is no rigorous theory of the dynamical behaviour of the vortex liquid in oscillating superflows. Experimentally, it may depend somewhat on the geometry of the oscillating body, in particular when tuning forks are considered, as will be shown below. The effect of remanent vorticity on the critical velocity at the first transition to turbulence after the measuring cell has been filled with helium will be discussed in terms of Kopnin's theory in Chapter \ref{experiments}.

\section{Universality}\label{model}

The dependence of $v_c \sim \sqrt{\kappa\,\omega}$ is a universal property of all oscillatory superflows and not restricted to $^4$He alone. To our knowledge, however, corresponding data on superfluid $^3$He are not yet conclusive (only 2 frequencies have been studied \cite{yano}) or, in case of Bose-Einstein condensates, are not yet available. Very recently, a universal scaling property of classical laminar oscillatory flows has been discovered by Ekinci {\it et al.}\cite{Ekinci}. These authors find that a scaling function exists that depends only on the dimensionless product of the oscillation frequency $\omega$ and the relaxation time $\tau$ of the liquid. While this scaling applies to oscillating classical laminar flows, at present no extension to oscillating superflows is available. What we have found \cite{jltp} is that quantum turbulence occurs only when $\omega \tau < 1/4$ where in our case $\tau = 2\kappa/\beta v_s^2$ is the relaxation time of the vortex liquid in a superfluid flow field $v_s$ as introduced by Kopnin \cite{kolya,jltp}. Similar to the situation in classical flows \cite{Ekinci} geometry and dimensions bear no effect. One might ask whether there is any relation between these cases. Therefore, it is desirable to have a rigorous theory of the dynamical behaviour of oscillating superflows.

An interesting consequence of the frequency dependence of $v_c$ is that the oscillation amplitude at the critical velocity $a_c = v_c/\omega \sim \sqrt{\kappa/\omega}$ is equal to the average vortex spacing $l_c = 1/\sqrt{L_c}$, where the vortex line length per unit volume is given by \cite{kolya} $L_c = (v_c/\kappa)^2$. This is a very plausible result. Finally, we note that only the oscillation amplitude determines the transition to turbulence and not the classical ``Strouhal number" $a_c/R$ which obviously has no significance for the onset of quantum turbulence \cite{oscflow}.

\section{Remanent vorticity}\label{experiments}

In the experiments both with spheres \cite{PRL,jltp} and, in particular, with wires \cite{yano} the critical velocity $v_c$ could be enormously exceeded in the first up-sweep of the oscillation amplitude and $v_c$ could only be determined in the down-sweep after the transition to turbulence finally had occurred.  We are attributing this hysteresis to a lack of remanent vorticity at the beginning of the experiment. In fact, detailed experiments by the Osaka group \cite{YanoJltp} demonstrate that remanent vorticity can be reduced or even avoided if the measuring cell is filled very slowly at low temperatures. In that case no transition to turbulence was observable up to very high velocity amplitudes of 1.5 m/s. Starting our analysis with an initial remanent vortex line length $L_0 \ll L_c$ we find from Kopnin's work \cite{kolya} that in this case the relaxation time is enhanced by a factor $(L_c/L_0)^{1/2}$. 
(No specific assumptions are made where and how these vortices are distributed throughout the measuring cell.) From the above condition for turbulence we now obtain a larger critical velocity $v\,^{\prime}_c \sim \omega \cdot l_0$\,where $l_0 > l_c$ is the average vortex spacing of the remanent vorticity $L_0$ that can be determined directly from the measured $v\,^{\prime}_c$ that is larger than $v_c$ by a factor $l_0/l_c = (L_c/L_0)^{1/2}$.
The critical velocity $v\,^{\prime}_c$ is reached when the oscillation amplitude $a\,^{\prime}_c = v\,^{\prime}_c/\omega$ is comparable to $l_0\,$: $a\,^{\prime}_c \sim l_0$, which is analogous to the result of Ch.2 where $a_c \sim l_c$ which means that in either case vortex lines must be within the reach of the oscillating body for turbulence to develop. This picture is qualitatively supported by numerical simulations \cite{HTV}.

\section{The numerical prefactor}

While the experimental results obtained with spheres, wires, and also a vibrating grid \cite{gridneu} follow the above behaviour, the data on tuning forks \cite{skrbek} seem to deviate towards values of $v_c$ lower than expected.
This discrepancy may be attributed to a geometrical effect on the prefactor of $v_c$. In general, the velocity amplitude of the flowing superfluid varies over the surface of the oscillating body. For spheres, {\it e.g.}, the flow velocity is largest at the equator, where it is increased by a factor of 1.5 compared to the velocity amplitude of the sphere. It appears plausible that turbulence originates predominantly in this region. For wires the corresponding increase is by a factor of 2.0. Experimentally, for spheres a prefactor of 2.8 is observed \cite{archiv} while for the most detailed data on wires obtained by the Osaka group \cite{yano} the prefactor is 2.1. This is expected from the ratio 1.5/2.0 = 0.75 = 2.1/2.8 but the perfect agreement may be accidental because there is still scatter of the data. 

For tuning forks the situation is more complicated. Both prongs of the fork are oscillating towards and away from each other. The liquid in between is pushed away or sucked in at a speed that will depend on the ratio of the width $W$ of the prongs to the spacing $D$. It is clear that for a ratio $W/D>1$ the flow velocity is enhanced by this factor. Moreover, because of the rectangular cross section of the prongs, the flow velocity will be enhanced even further near the edges. Hence, the measured values of $v_c$ will be lower and will depend on the geometry of the forks. Experimentally, there is some evidence for that but it is rather qualitative because it is impossible to vary the resonance frequency of a fork without changing the geometry as well \cite{skrbek}.

\section{Crossover to steady flows}

\begin{figure}
\centerline{
\includegraphics[width=0.9\linewidth,clip=true]{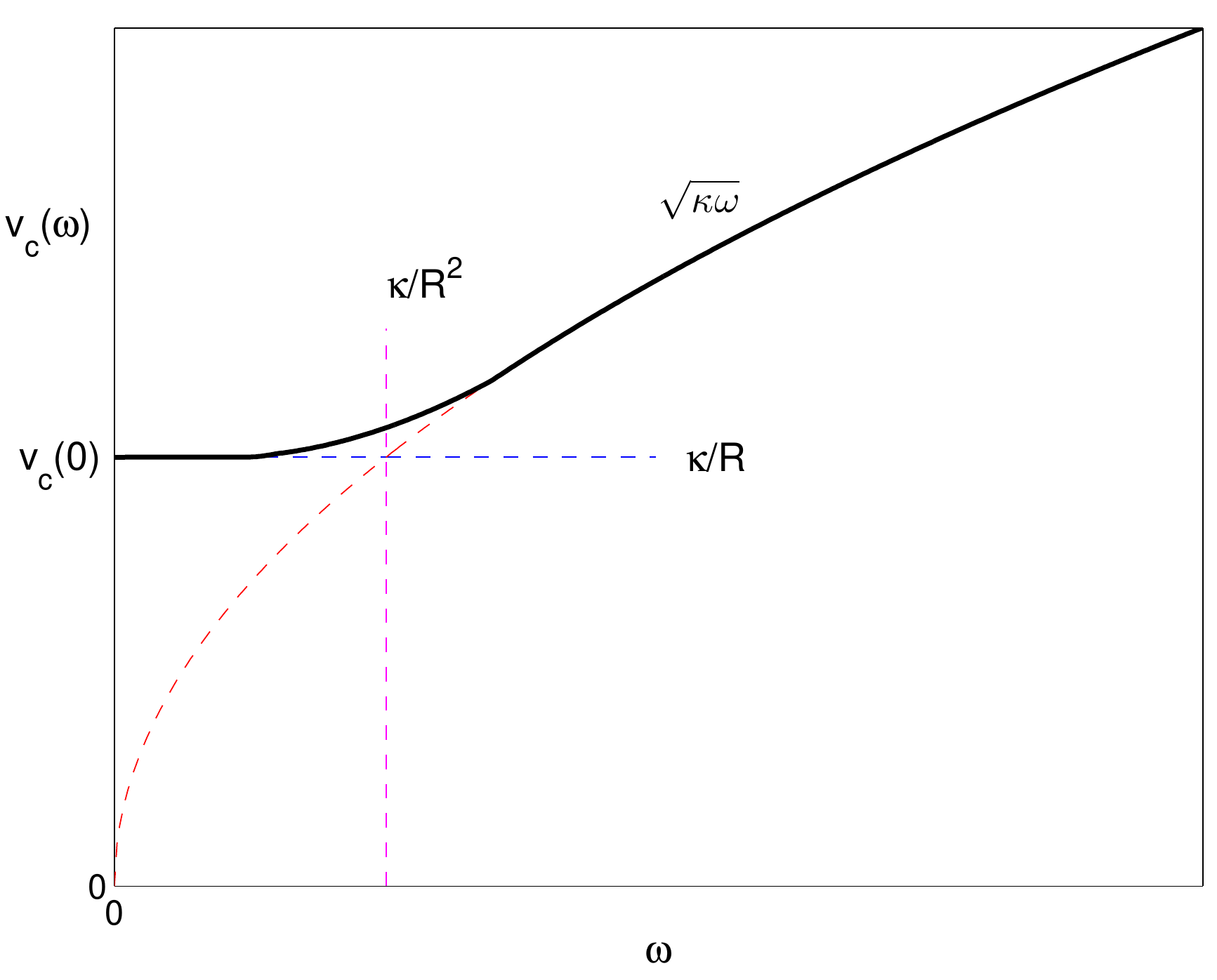}
}
\caption{(Color online) Crossover of the critical velocity $v_c(\omega)$ from oscillatory flow to steady flow $v_c(0)$ at $\omega \sim \kappa/R^2$, where $R$ is
 the size of the object. 
}
\label{f.vc}
\end{figure}

The question arises how the frequency dependence of $v_c(\omega)$ connects to the steady case where $\omega = 0$. Of course, a finite critical velocity is known to exist in that case too. Clearly, there must be a crossover to the steady case before $\omega$ goes to zero. We again consider the superfluid Reynolds number $Re_s = vl/\kappa$. At low frequencies the characteristic length scale $l$ must now be the size $R$ of the object because the oscillation amplitude $a$ diverges: $R \ll a$. This gives then $v_c(0) \sim \kappa/R$, which one would also expect on dimensional grounds \cite{archiv} and which is well known from vortex ring production in steady flows (except for a factor $1/2\pi$ and a logarithmic correction). Therefore, the crossover occurs when $\omega \sim  \kappa/ R^2$ (see Fig.1), {\it i.e.}, when the average vortex separation $l_c$ at the critical velocity or, equivalently, the critical oscillation amplitude $a_c \sim l_c \sim \sqrt{\kappa/ \omega}$ begin to exceed $R$. For our spheres of size (diameter) $R = 0.2$ mm we obtain $v_c(0) \sim 0.5$ mm/s and a crossover at $\omega \sim 2.5$ s$^{-1}$ which is much smaller than the oscillation frequencies of our spheres $\ge$ 750 s$^{-1}$ and, therefore, could not be observed in our experiments. For wires, however, having a size of only 3 micrometer \cite{yano} the transition will occur already at $1.1\cdot10^4$ s$^{-1}$, or 1.8 kHz. For frequencies that are substantially lower, the critical velocities will reach the constant level of $v_c(0) \sim \kappa/R = 33$ mm/s (when numerical prefactors are taken into account these numbers will change by a factor of order 1).

\begin{acknowledgments}
We appreciate valuable discussion with N.B. Kopnin, B.V. Svistunov, G.E. Volovik, and H. Yano who also showed us his experimental results before publication. R.H. acknowledges the support from the Academy of Finland (Grant 114887).
\end{acknowledgments}



\end{document}